\documentclass[fleqn,usenatbib]{mnras}

\usepackage[T1]{fontenc}
\usepackage{ae,aecompl}

\usepackage{amssymb,amsmath,amsfonts,graphicx,epsf}

\title{Influence of a large-scale field on energy dissipation in magnetohydrodynamic turbulence}

\author[V. Zhdankin et al.]{
Vladimir Zhdankin,$^{1}$\thanks{E-mail: zhdankin@jila.colorado.edu}
Stanislav Boldyrev,$^{2,3}$
Joanne Mason$^{4}$
\\
$^{1}$JILA, NIST and University of Colorado, 440 UCB, Boulder, Colorado 80309, USA\\
$^{2}$Department of Physics, University of Wisconsin-Madison, 1150 University Avenue, Madison, Wisconsin 53706, USA\\
$^{3}$Space Science Institute, Boulder, Colorado 80301, USA\\
$^{4}$College of Engineering, Mathematics \& Physical Sciences, University of Exeter, North Park Road, Exeter EX4 4QF, UK\\
}

\date{Accepted XXX. Received YYY; in original form ZZZ}

\pubyear{2016}

\begin{document}
\label{firstpage}
\pagerange{\pageref{firstpage}--\pageref{lastpage}}
\maketitle

\begin{abstract}
In magnetohydrodynamic (MHD) turbulence, the large-scale magnetic field sets a preferred local direction for the small-scale dynamics, altering the statistics of turbulence from the isotropic case. This happens even in the absence of a total magnetic flux, since MHD turbulence forms randomly oriented large-scale domains of strong magnetic field. It is therefore customary to study small-scale magnetic plasma turbulence by assuming a strong background magnetic field relative to the turbulent fluctuations. This is done, for example,  in reduced models of plasmas, such as reduced MHD, reduced-dimension kinetic models, gyrokinetics, etc., which make theoretical calculations easier and numerical computations cheaper. Recently, however, it has become clear that the turbulent energy dissipation is concentrated in the regions of strong magnetic field variations. A significant fraction of the energy dissipation may be localized in very small volumes corresponding to the boundaries between strongly magnetized domains. In these regions the reduced models are not applicable. This has important implications for studies of particle heating and acceleration in magnetic plasma turbulence. The goal of this work is to systematically investigate the relationship between local magnetic field variations and magnetic energy dissipation, and to understand its implications for modeling energy dissipation in realistic turbulent plasmas.
\end{abstract}




\section{Introduction}

Large-scale magnetic fields are an essential part of magnetohydrodynamic (MHD) turbulence. Even when a large-scale magnetic field is not imposed on the system externally (as may be done in laboratory devices, for example), it is generated by turbulence due to magnetic dynamo action. Such a magnetic field plays a crucial role in  magnetic turbulence at small scales. Indeed, unlike a uniform large-scale velocity field, the large-scale magnetic field cannot be removed by a Galilean transformation, and it mediates the energy cascade at all scales. Turbulent plasmas with high Reynolds numbers are therefore anisotropic at small scales with the eddies stretched along a local strong background field. This allows various reduced models, such as reduced MHD models, gyrokinetics models, or models with reduced spatial dimensionality \citep[e.g.,][]{dmitruk_etal_2005, perez2008, tobias2011, mason2012, schekochihin_etal_2009, camporeale2011,wu2013,karimabadi2013,franci2015}, to accurately describe the local dynamics. Recently, there has been a widespread application of these reduced models to understand the dissipation of the turbulent cascade in the solar wind \citep{camporeale2011,wu2013,karimabadi2013,franci2015, howes_etal_2008, howes_etal_2008b, boldyrev2011, howes_etal_2011, tenbarge_howes_2013, tenbarge_etal_2013, told_etal_2015} and in the solar corona \citep{einaudi1999, oughton_etal_2001, dmitruk_etal_2002, rappazzo_etal_2007, rappazzo_etal_2008, wan_etal_2014}.

On the other hand, a large fraction of energy is now known to be dissipated in a small fraction of the volume that is often characterized by strong variations in the large-scale magnetic field  \citep[e.g.,][ and references therein]{zhdankin_etal_2016b}. In fact, one may think of MHD turbulence without an imposed large-scale field as consisting of subdomains where the large-scale magnetic field is strong, separated by thin boundaries where the direction of large-scale field changes abruptly. This structure is consistent, for example, with the Borovsky picture of solar wind turbulence as an ensemble of tightly packed flux tubes \cite[][]{borovsky2008}, and with statistical significance of strong rotational magnetic discontinuities observed in other studies~\cite[e.g.,][]{bruno_etal_2001, li2008, zhdankin_etal_2012, greco_etal_2009}.  The regions of strong magnetic field variations are very intermittent, i.e., they occupy a small volume and contain only a small fraction of the total energy. However, they may contain a significant fraction of the magnetic energy dissipation. If so, the reduced model may not be used to properly describe turbulent energy dissipation. The question of the extent to which energy dissipation is skewed toward the regions of strong variation of the magnetic field is therefore of principal importance for phenomenological and numerical modeling of MHD turbulence. This work presents the quantitative statistical analysis of this issue. 

In this work, we utilize numerical simulations of MHD turbulence to investigate the local relationship between the energy dissipation rate and the relative variation of the magnetic field.  We find that a significant fraction of the energy dissipation occurs in regions where this variation is large, although this fraction slowly decreases with increasing size of the inertial range. We therefore argue that caution is required when applying reduced models to systems where the inertial interval is not sufficiently long. For instance, we estimate that in systems where the inertial interval spans less than three orders of magnitude, more than 15\% of the energy dissipation occurs in the regions with strong variations of the magnetic field, where the reduced models are not applicable. These regions occupy a very small fraction of the volume, however, the dissipation inside them is very intense. Models that fail to properly account for such regions may lead to an incomplete description of energy dissipation and particle heating in MHD plasma turbulence. 

\section{Methods}

For our analysis we numerically solve the incompressible MHD equations for the plasma velocity $\boldsymbol{v}(\boldsymbol{x},t)$ and the magnetic field $\boldsymbol{B}(\boldsymbol{x},t)=\boldsymbol{B}_0+\boldsymbol{b}(\boldsymbol{x},t)$ (where $\boldsymbol{B}_0 = B_0 \hat{\boldsymbol{z}}$ is the uniform background field):
\begin{eqnarray}
& \partial_t \boldsymbol{v} + (\boldsymbol{v} \cdot \nabla) \boldsymbol{v} = - \nabla p + (\nabla \times \boldsymbol{B}) \times \boldsymbol{B} + \nu \nabla^2 \boldsymbol{v} + \boldsymbol{f}_1, \nonumber \\
& \partial_t \boldsymbol{B} = \nabla \times (\boldsymbol{v} \times \boldsymbol{B}) + \eta \nabla^2 \boldsymbol{B} +\boldsymbol{f}_2, \label{eq:mhd}
\end{eqnarray}
where $p$ is the plasma pressure, along with $\nabla \cdot \boldsymbol{v} = \nabla \cdot \boldsymbol{B} = 0$. For simplicity, we take the resistivity to be equal to the viscosity, $\eta = \nu$, in the simulations. The turbulence is driven by random large-scale external forces $\boldsymbol{f_1}$ and $\boldsymbol{f_2}$ that are applied in Fourier space and have amplitudes chosen so that $v_{rms}\approx 1$. The forces have no component in the direction of the background field and are solenoidal in the $xy$-plane. The Fourier coefficients are non-zero only for wavenumbers $k_{x,y,z}=\pm 1$ or $ \pm 2$, and in such cases the coefficients are chosen from a Gaussian distribution and are refreshed on average every $0.1L/(2\pi v_{rms} )$ time units, where $L$ is the size of the domain (that is, the force is updated approximately 10 times per large-scale turnover time). While the particular choice of the force's statistical properties and correlation time does not affect the spectra of turbulence at smaller scales \cite[][]{mason2006,mason_cb08}, this setup allows us to supply energy in large-scale Alfv\'enic fluctuations in a controlled fashion. The equations are solved on a triply periodic domain using standard pseudospectral methods. Time advancement of the diffusive terms is carried out exactly using the integrating factor method, while the remaining terms are treated using a third-order Runge-Kutta scheme. Further details of the numerical approach can be found in \cite{cattaneo_etal_2003}. A background magnetic field with relatively small amplitude $B_0 = 0.5 b_{\rm rms}$ is imposed. For the present analysis, we focus on a simulation with a $1024^3$ lattice and a Reynolds number $\operatorname{Re}= v_{\rm rms} (L/2\pi) / \nu \sim 5500$; simulations with smaller $\operatorname{Re}$ give similar results. We carry out our analysis on 5 snapshots, each separated by 2 eddy turnover times.

We aim to understand how the relative amplitude of the magnetic fluctuations is correlated with the local energy dissipation rate. For that we have designed the following statistical approach. We subdivide the simulation domain into cubes of size $\Delta x$ and consider the statistical properties of fluctuations in the cubes. Consider a cube of size $\Delta x$, which is centered at the point $\boldsymbol{x}$. The local mean magnetic field in this cube is then given by
\begin{align}
\bar{\boldsymbol{B}}_{\Delta x}(\boldsymbol{x}) = \frac{1}{(\Delta x)^3} \int\limits_{\Delta x} d^3x' \boldsymbol{B}(\boldsymbol{x} + \boldsymbol{x}') \, , 
\end{align}
while the root-mean-square (rms) magnetic field fluctuation is given by
\begin{align}
B_{\text{rms},\Delta x}(\boldsymbol{x}) = \left[ \frac{1}{(\Delta x)^3} \int\limits_{\Delta x} d^3x' \left|\boldsymbol{B}(\boldsymbol{x} + \boldsymbol{x}') - \bar{\boldsymbol{B}}_{\Delta x}(\boldsymbol{x})\right|^2 \right]^{1/2} \, ,
\end{align}
where $\int_{\Delta x}$ denotes an integral across the volume of the cube. The local strength of the fluctuations is then characterized by the ratio $R_{\Delta x} = B_{\text{rms},\Delta x}/\bar{B}_{\Delta x}$. The presence of a strong local magnetic field is implied by $R_{\Delta x} \ll 1$. The local energy dissipation rate in a cube centered at point $\boldsymbol{x}$ is given by
\begin{align}
{\mathcal E}_{\Delta x}(\boldsymbol{x}) &= \int\limits_{\Delta x} d^3x' \left[ \eta \left|\boldsymbol{j}(\boldsymbol{x} + \boldsymbol{x}')\right|^2 + 2 \nu {\bf \sigma}(\boldsymbol{x} + \boldsymbol{x}') : {\bf \sigma}(\boldsymbol{x} + \boldsymbol{x}') \right] \, ,
\end{align}
where $\boldsymbol{j} = \nabla \times \boldsymbol{B}$ is the current density and ${\bf \sigma} = [\nabla \boldsymbol{v} + (\nabla \boldsymbol{v})^T]/2$ is the rate-of-strain tensor. We note that ${\mathcal E}_{\Delta x}$ includes contributions from both resistive and viscous dissipation, but our results are broadly similar when either dissipation mechanism is considered individually. We measure the above quantities for cubes of varying size $\Delta x$ in order to understand the scale dependence of the field fluctuations and energy dissipation. Our major object of study is the correlation between the local intensity of fluctuations $R_{\Delta x}$ and the local energy dissipation rate ${\mathcal E}_{\Delta x}$.

\begin{figure}
\includegraphics[width=\columnwidth]{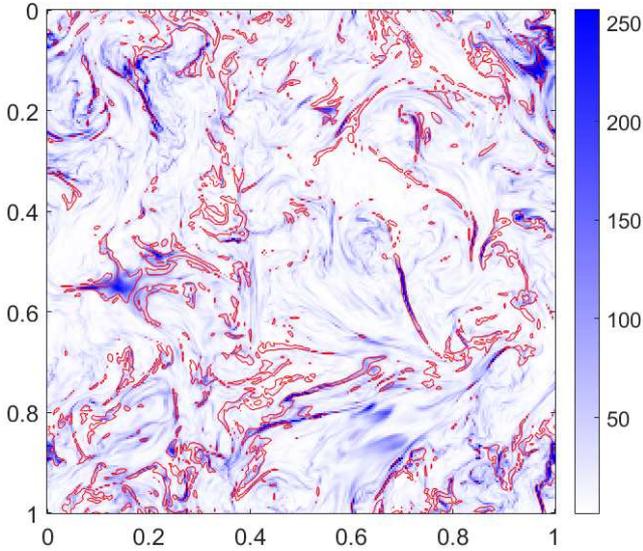}
 \centering
  \caption{\label{fig:profile} Contour plot of fluctuation-to-mean ratio $R_{\Delta x}$ (red) overlaid on an image of the local energy dissipation rate ${\mathcal E}_{\Delta x}$ (blue) for $\Delta x/L = 1/256$, in an $xy$ slice of the simulation. The contours are taken at $R_{\Delta x} = 1/3$, and the colorbar indicates $30 \times {\mathcal E}_{\Delta x} / \langle {\mathcal E}_{\Delta x} \rangle$. A strong correlation between the magnetic field variations and the energy dissipation is observed.}
 \end{figure}

For reference, in Fig.~\ref{fig:profile} we show contours of $R_{\Delta x} = 1/3$ overlaid on an image of the local energy dissipation rate ${\mathcal E}_{\Delta x}$ in an arbitrarily chosen 2D plane of the simulation, for $\Delta x/L = 1/256$. There is evidently a strong degree of correlation between the two quantities, with both often being concentrated in thin, sheet-like coherent structures. 

  \section{Results}
 
\subsection{The mean field and the fluctuations}

\begin{figure}
\includegraphics[width=\columnwidth]{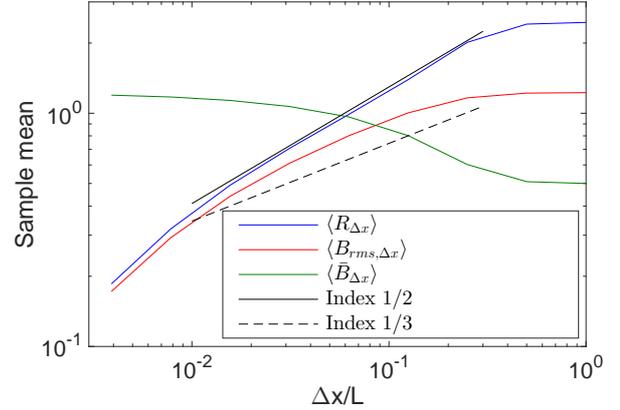}
 \centering
  \caption{\label{fig:unconditioned} The mean ratio $\langle R_{\Delta x} \rangle = \langle B_{{\rm rms},\Delta x}/ \bar{B}_{\Delta x} \rangle$ (blue), mean fluctuations $\langle B_{{\rm rms},\Delta x} \rangle$ (red), and mean field $\langle \bar{B}_{\Delta x} \rangle$ (green) versus scale $\Delta x$. The power-law scalings $\Delta{x}^{1/2}$ (black, solid) and $\Delta{x}^{1/3}$ (black, dashed) are shown for reference.}
 \end{figure}

We begin by analysing the statistical properties of the {\em mean} local quantities $\langle \bar{B}_{\Delta x} \rangle$, $\langle B_{\rm{rms}, \Delta x} \rangle$ and $\langle R_{\Delta x} \rangle$, where the angular brackets denote averaging over all the cubes of size $\Delta x$ in the simulation domain. Fig.~\ref{fig:unconditioned} shows the scaling of these quantities versus $\Delta x$. We see that the local mean field measured at progressively smaller scales, $\langle \bar{B}_{\Delta x} \rangle\vert_{\Delta x\to 0}$, approaches the value of the large-scale fluctuations, $\langle B_{{\rm rms},\Delta x} \rangle\vert_{\Delta x\to L}$, confirming that the large-scale magnetic field fluctuations act as a local background field for the small-scale fluctuations. This important fact, quantified in Fig.~\ref{fig:unconditioned}, is behind the applicability of models of MHD turbulence that assume a strong imposed uniform magnetic field. 

In addition, we see that, to a very good approximation, $\langle R_{\Delta x} \rangle \propto (\Delta x)^{1/2}$ in the inertial range, while the fluctuations $\langle B_{{\rm rms},\Delta x} \rangle$ and mean $\langle \bar{B}_{\Delta x} \rangle$ are not as well fit by power laws. One may, however, roughly approximate $\langle B_{{\rm rms},\Delta x} \rangle \propto (\Delta x)^{1/3}$, which is broadly consistent with the scaling ($1/3$) of magnetic field increments in the Goldreich-Sridhar phenomenology \cite[][]{goldreich1995}, and not far from the scaling ($1/4$) predicted in the model of scale-dependent dynamic alignment \cite[e.g.,][]{boldyrev2006,perez_mason2012}. We however note that these phenomenologies assume the presence of a {\em strong} and {\em constant} large-scale magnetic field in the whole domain, whereas in our measurement we instead average over the cubes with all possible values of the mean field.

Finally, we note that the observed scaling of the fluctuations-to-mean ratio, $\langle R_{\Delta x} \rangle \propto (\Delta x)^{1/2}$, implies that anisotropy grows significantly with decreasing scale.  For example, $\langle R_{\Delta x} \rangle$ decreases from $1$ to roughly~$1/10$ after $\Delta x$ decreases by only two decades. This implies that the reduced models of MHD turbulence should formally be valid for a description of energy distribution in the bulk of the small-scale fluctuations in most space and astrophysical systems. 

\subsection{The energy dissipation}

The picture changes significantly when we consider the {\em energy dissipation}, which is known to be very intermittent, that is, not space filling \cite[e.g.,][]{osman2012,zhdankin_etal_2016b}. We now analyze the correlation of the local energy dissipation rate ${\mathcal E}_{\Delta x}$ with $R_{\Delta x}$.

\begin{figure}
\includegraphics[width=\columnwidth]{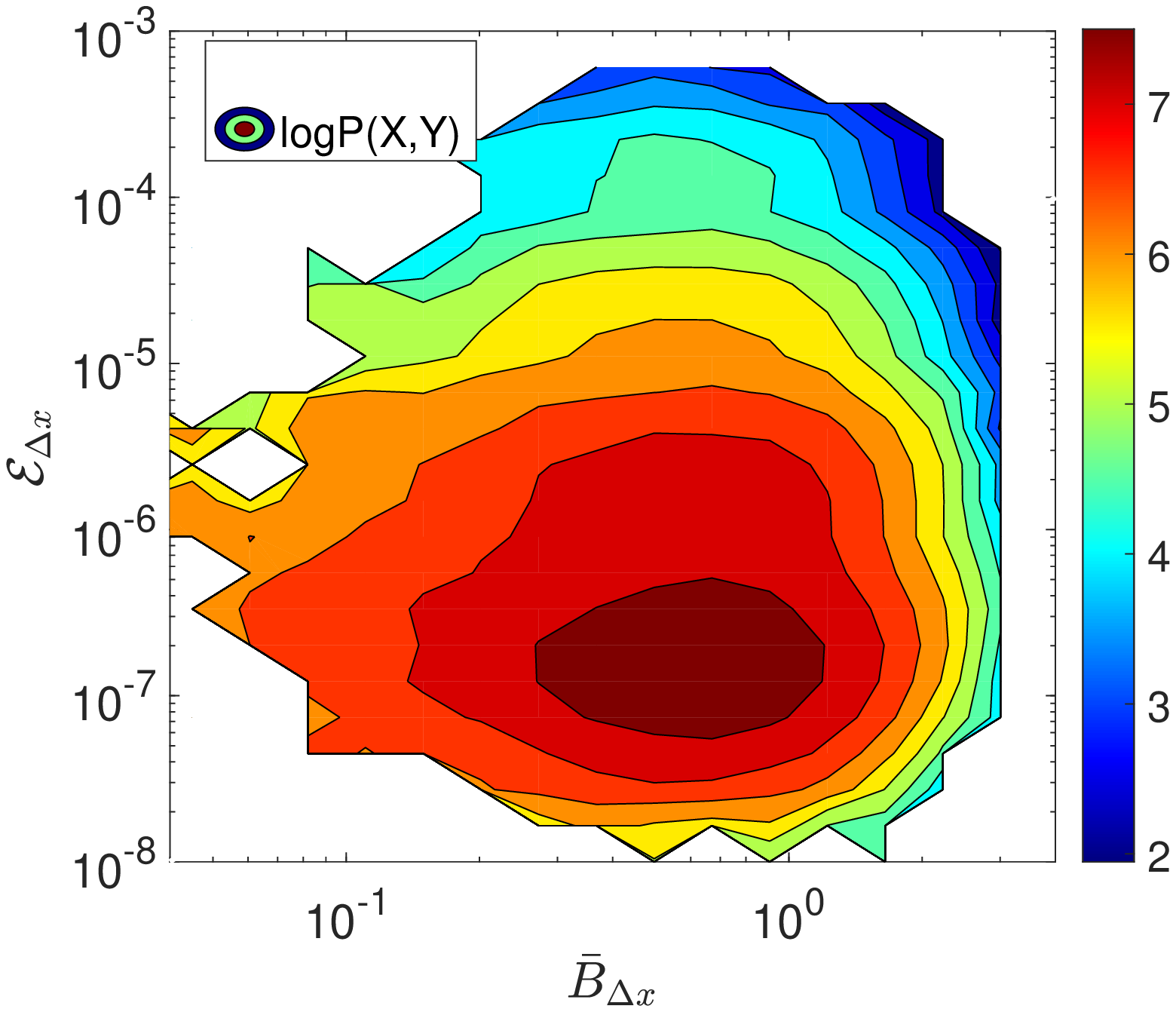}
\includegraphics[width=\columnwidth]{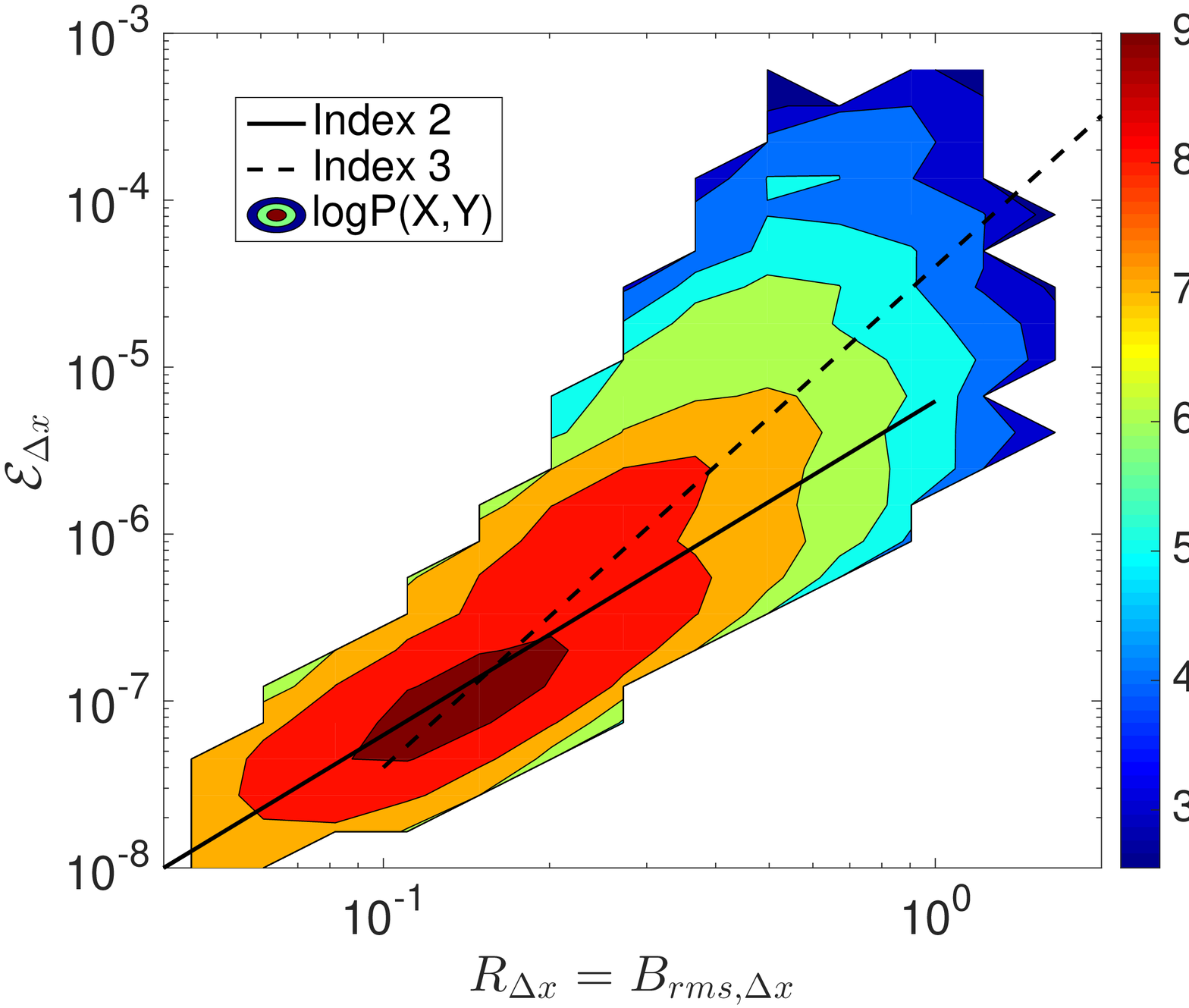}
\includegraphics[width=\columnwidth]{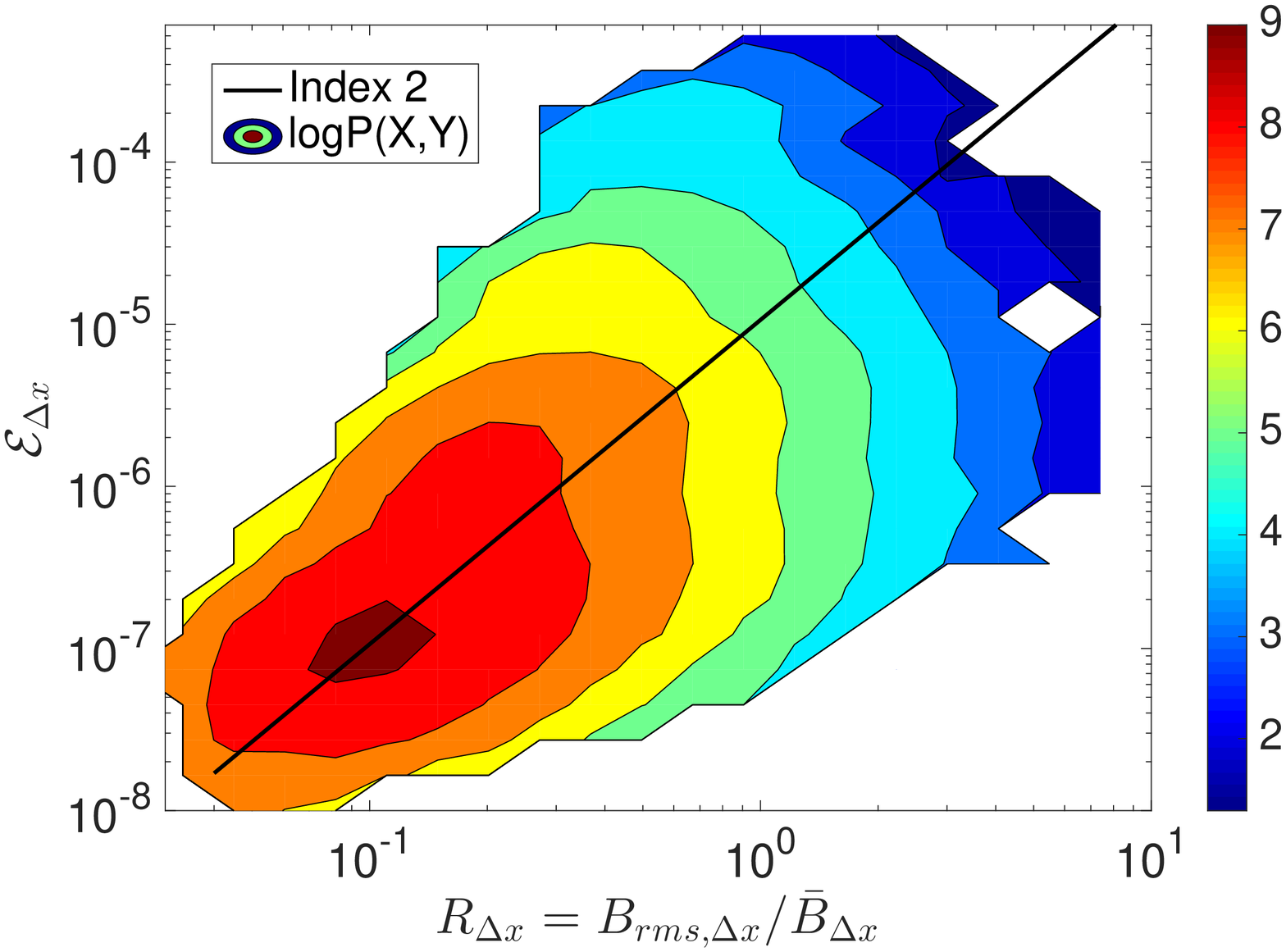}
 \centering
  \caption{\label{fig:correlation} The 2D joint probability density function of the local energy dissipation rate ${\mathcal E}_{\Delta x}$ and mean magnetic field $\bar{B}_{\Delta x}$ (top panel), rms fluctuations $B_{\text{rms},\Delta x}$ (center panel), and rms-to-mean ratio $R_{\Delta x} = B_{\text{rms},\Delta x}/\bar{B}_{\Delta x}$ (bottom panel), for $\Delta x/L = 1/128$ (the plots for other $\Delta x$ are similar). A quadratic scaling (solid black line) and cubic scaling (dashed black line) are shown for reference.}
 \end{figure}

In Fig.~\ref{fig:correlation}, we show 2D joint probability density functions of ${\mathcal E}_{\Delta x}$ and each of $\bar{B}_{\Delta x}$, $B_{{\rm rms},\Delta x}$, and $R_{\Delta x}$ separately. We find a strong correlation between the dissipation and the fluctuations, such that the most intense dissipation indeed takes place in regions of large relative fluctuations in the magnetic field. In particular, we find that the results can be fit rather well by a quadratic scaling, ${\mathcal E}_{\Delta x} \propto R_{\Delta x}^2$. The scaling of dissipation with absolute fluctuations can be approximated by ${\mathcal E}_{\Delta x} \propto B_{{\rm rms},\Delta x}^3$ for inertial-range fluctuations and ${\mathcal E}_{\Delta x} \propto B_{{\rm rms},\Delta x}^2$ for weaker fluctuations. On the other hand, there is very little correlation between the dissipation and local mean field $\bar{B}_{\Delta x}$, consistent with the mean field being set by the background, large-scale eddies.

While the scaling of ${\mathcal E}_{\Delta x}$ with $R_{\Delta x}$ is non-trivial to explain, we note that a cubic scaling of the dissipation with respect to $B_{{\rm rms},\Delta x}$ can be expected on general grounds. This is because the local dissipation should be comparable to the local magnetic energy divided by the cascade time, which can be estimated by the local eddy turnover time, $\tau_{\Delta x} \sim \Delta x / B_{{\rm rms},\Delta x}$ (in simulation units). We then arrive at
\begin{align}
{\mathcal E}_{\Delta x} \sim (\Delta{x})^3 \frac{B_{{\rm rms},\Delta x}^2}{\tau_{\Delta x}} \sim (\Delta{x})^2  B_{{\rm rms},\Delta x}^3 \, . \label{eq:corr}
\end{align}

\begin{figure}
\includegraphics[width=\columnwidth]{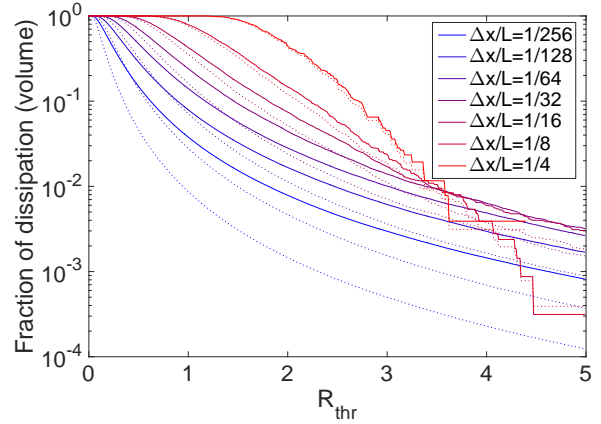}
\caption{\label{fig:dist_cum_1} The fraction of total energy dissipation occurring in cubes of size $\Delta x$ with the ratio of local fluctuations-to-mean exceeding a threshold, $R_{\Delta x} > R_{\rm thr}$, for $\Delta x/L \in \{ 1/256, 1/128, 1/64, 1/32, 1/16, 1/8, 1/4 \}$ (blue to red, solid lines). The corresponding fraction of volume occupied by the cubes is also shown (dotted lines).}
\end{figure}

To assess more quantitatively the extent to which dissipation occurs in regions with the large fluctuations-to-mean ratio, $R_{\Delta x} \gtrsim 1$, we consider the cumulative distribution of energy dissipation conditioned on $R_{\Delta x}$, which we denote by ${\mathcal E}_{{\rm cum}, R_{\Delta x}} (R_{\rm thr})$. In particular, we set a threshold $R_{\rm thr}$ and measure the fraction of the total energy dissipation that occurs in cubes with $R_{\Delta x} > R_{\rm thr}$. The results are shown in Fig.~\ref{fig:dist_cum_1}, along with the volume occupied by cubes exceeding the threshold. The cumulative distributions extend to large values of $R_{\Delta x}$, implying that, indeed,  a significant fraction of energy dissipation may occur in regions with $R_{\Delta x} \gtrsim 1$. However, the tail of the distribution  function shifts downwards for decreasing $\Delta x$. Hence, at sufficiently small scales, the majority of energy dissipation should occur in regions where $R_{\Delta x}$ is small. This means that asymptotically in the limit of very large Reynolds number, both the small-scale fluctuations and the energy dissipation are adequately captured by the reduced models.   

\begin{figure}
\includegraphics[width=\columnwidth]{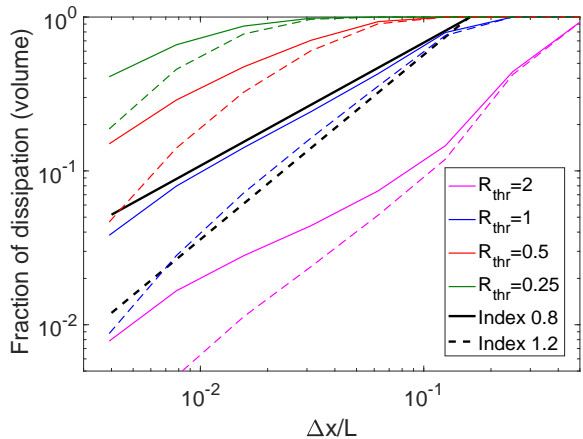}
\caption{\label{fig:dist_cum} The fraction of total energy dissipated (solid colored lines) and volume occupied (dashed colored lines) in cubes of size $\Delta x$ with the ratio of local fluctuations-to-mean exceeding a threshold, $R_{\Delta x} > R_{\rm thr}$, for $R_{\rm thr} = 2$ (magenta), $1$ (blue), $1/2$ (red), and $1/4$ (green). Power-law scalings $\Delta x^{0.8}$ (solid black line) and $\Delta x^{1.2}$ (dashed black line) are shown for reference (black).}
 \end{figure}

As we now demonstrate, however, this convergence is rather slow. Figure~\ref{fig:dist_cum} shows the dependence of the cumulative energy dissipation ${\mathcal E}_{{\rm cum}, R_{\Delta x}}(R_{\rm thr})$ on the scale $\Delta x$, for several values of $R_{\rm thr}$. From the plot corresponding to $R_{\rm thr}=1$ we estimate a scaling ${\mathcal E}_{{\rm cum}, R_{\Delta x}} (1) \propto (\Delta x)^{0.8}$. The curves with $R_{\rm thr} < 1$ are similar; they seem to have the same scaling but shifted upward with respect to the curve with $R_{\rm thr} = 1$. To understand the implications of this slow convergence we analyze the following example. Consider the curve corresponding to the threshold $R_{\rm thr} = 1/4$. Such a threshold approximately separates the cubes where the reduced MHD model provides a good description for the spectrum of MHD turbulence ($R_{\Delta x}<1/4$) from the cubes where it does not ($R_{\Delta x}>1/4$) \cite[e.g.,][]{mason2006,mason_cb08,mason2012}. Assuming that we may extrapolate the observed scaling to very small $\Delta x$, we can estimate for this curve:  ${\mathcal E}_{{\rm cum}, R_{\Delta x}} (1/4) \sim 40 \times (\Delta x/L)^{0.8}$. The fraction of the dissipation occurring inside the cubes with $R_{\Delta x}<1/4$ will thus exceed $85\%$ if $\Delta x/L < 10^{-3}$, and $97\%$ if $\Delta x/L < 10^{-4}$. This means, for example, that the reduced MHD model will correctly capture more than $97\%$ of the energy dissipation only if the inertial interval of the turbulence extends to scales smaller than $\Delta x/L < 10^{-4}$. In this case, less than $3\%$ of the magnetic energy dissipation will happen inside the cubes where magnetic fluctuation-to-mean ratio exceeds $1/4$, that is, where the reduced models are not applicable. 

The regions with $R_{\rm thr} > 1/4$, corresponding to $3\%$ of all the energy dissipation in the considered  example, are, however, extremely intense. From Fig.~\ref{fig:dist_cum} we estimate the volume occupied by the structures with $R_{\rm thr} > 1$ as $V_{R_{\Delta x}}(1)\propto (\Delta x)^{1.2}$. Assuming that the same scaling holds for smaller values of $R_{\rm thr}$, we can estimate from Fig.~\ref{fig:dist_cum} that $V_{R_{\Delta x}}(1/4)\sim 150\times (\Delta x/L)^{1.2}$. The cubes of the size $\Delta x/L=10^{-4}$, which correspond to $R_{\rm thr} > 1/4$, considered in the previous example, will therefore occupy only about 0.2\% of the total volume. They include significant variations in the magnetic field direction. In cases where the energy dissipation or particle acceleration effects are strongly skewed toward the environments with strong variations of the magnetic field direction \cite[e.g.,][]{chen2015,chasapis2015,tessein2016}, these effects will not be adequately captured by the reduced models. 

\section{Conclusions}

Magnetic plasma turbulence is intrinsically anisotropic, meaning that small scale fluctuations experience a large-scale magnetic field that mediates nonlinear interactions. This happens even if the strong magnetic field is not imposed externally, since such a field is self-consistently generated by turbulence itself. It is therefore customary in studies of MHD turbulence to assume the presence of a strong uniform background magnetic field. This assumption, introduced mostly phenomenologically in previous studies, is put on a firmer, quantitative ground in our work. In particular, we argued that the so-called reduced models, that is, models assuming a strong background field and correspondingly anisotropic fluctuations (reduced MHD, gyrokinetics, models with reduced dimensionality, etc.), should describe the turbulent energy distribution correctly, when their inertial interval is sufficiently long. We however also established that such models may not work as well for describing the energy dissipation in systems where the inertial interval for magnetic fluctuations is not long enough. The reason is that the dissipation is extremely spatially intermittent. It is skewed toward the regions where the magnetic field fluctuations are relatively large compared to the mean field, as happens at the boundaries between nearly uniformly magnetized domains. For example, we estimated that more than 3\% of the energy dissipation is not captured by the reduced models if the MHD inertial interval extends over less than four orders of magnitude.  Such constraints may be relevant for some natural systems (e.g., solar wind turbulence \cite[][]{kiyani_etal_2015}), and they may also be essential for laboratory experiments, say liquid metal experiments, where the magnetic Reynolds numbers are not large enough \cite[e.g.,][]{lathrop2005}. Moreover, the regions of strong energy dissipation occupy very small volumes, and, therefore, they may be extremely intense. No matter how large the Reynolds number is, the reduced models always miss a certain fraction of intense dissipation events generated by turbulence, which may become important in phenomena involving higher-order moments of field variations, say, transport phenomena. 

\section*{Acknowledgements}

The authors thank the referee, Alexander Schekochihin, for helpful comments. VZ acknowledges support from NSF grant AST-1411879. SB is partly supported by the National Science Foundation under the grant NSF AGS-1261659 and by the Vilas Associates Award from the University of Wisconsin - Madison. JM acknowledges the support of the EPSRC, through grant EP/M004546/1. We acknowledge PRACE for awarding us access to resource FERMI based in Italy at CINECA, and the STFC DiRAC HPC Facility for access to the COSMA Data Centric system at Durham University and MINERVA at the University of Warwick.




\bibliographystyle{mnras}
\bibliography{refs_all} 

\begin{thebibliography}{}
\makeatletter
\relax
\def\mn@urlcharsother{\let\do\@makeother \do\$\do\&\do\#\do\^\do\_\do\%\do\~}
\def\mn@doi{\begingroup\mn@urlcharsother \@ifnextchar [ {\mn@doi@}
  {\mn@doi@[]}}
\def\mn@doi@[#1]#2{\def\@tempa{#1}\ifx\@tempa\@empty \href
  {http://dx.doi.org/#2} {doi:#2}\else \href {http://dx.doi.org/#2} {#1}\fi
  \endgroup}
\def\mn@eprint#1#2{\mn@eprint@#1:#2::\@nil}
\def\mn@eprint@arXiv#1{\href {http://arxiv.org/abs/#1} {{\tt arXiv:#1}}}
\def\mn@eprint@dblp#1{\href {http://dblp.uni-trier.de/rec/bibtex/#1.xml}
  {dblp:#1}}
\def\mn@eprint@#1:#2:#3:#4\@nil{\def\@tempa {#1}\def\@tempb {#2}\def\@tempc
  {#3}\ifx \@tempc \@empty \let \@tempc \@tempb \let \@tempb \@tempa \fi \ifx
  \@tempb \@empty \def\@tempb {arXiv}\fi \@ifundefined
  {mn@eprint@\@tempb}{\@tempb:\@tempc}{\expandafter \expandafter \csname
  mn@eprint@\@tempb\endcsname \expandafter{\@tempc}}}

\bibitem[\protect\citeauthoryear{Boldyrev}{Boldyrev}{2006}]{boldyrev2006}
Boldyrev S.,  2006, Physical Review Letters, 96, 115002

\bibitem[\protect\citeauthoryear{Boldyrev, Perez, Borovsky  \&
  Podesta}{Boldyrev et~al.}{2011}]{boldyrev2011}
Boldyrev S.,  Perez J.,  Borovsky J.,   Podesta J.,  2011, The Astrophysical
  Journal Letters, 741, L19

\bibitem[\protect\citeauthoryear{Borovsky}{Borovsky}{2008}]{borovsky2008}
Borovsky J.,  2008, Journal of Geophysical Research, 113, A08110

\bibitem[\protect\citeauthoryear{Bruno, Carbone, Veltri, Pietropaolo  \&
  Bavassano}{Bruno et~al.}{2001}]{bruno_etal_2001}
Bruno R.,  Carbone V.,  Veltri P.,  Pietropaolo E.,   Bavassano B.,  2001,
  Planetary and Space Science, 49, 1201

\bibitem[\protect\citeauthoryear{{Camporeale} \& {Burgess}}{{Camporeale} \&
  {Burgess}}{2011}]{camporeale2011}
{Camporeale} E.,  {Burgess} D.,  2011, \mn@doi [\apj]
  {10.1088/0004-637X/730/2/114}, \href
  {http://adsabs.harvard.edu/abs/2011ApJ...730..114C} {730, 114}

\bibitem[\protect\citeauthoryear{Cattaneo, Emonet  \& Weiss}{Cattaneo
  et~al.}{2003}]{cattaneo_etal_2003}
Cattaneo F.,  Emonet T.,   Weiss N.,  2003, The Astrophysical Journal, 588,
  1183

\bibitem[\protect\citeauthoryear{{Chasapis} et~al.,}{{Chasapis}
  et~al.}{2015}]{chasapis2015}
{Chasapis} A.,  et~al., 2015, \mn@doi [Astrophys. J. Lett.]
  {10.1088/2041-8205/804/1/L1}, \href
  {http://adsabs.harvard.edu/abs/2015ApJ...804L...1C} {804, L1}

\bibitem[\protect\citeauthoryear{{Chen}, {Matteini}, {Burgess}  \&
  {Horbury}}{{Chen} et~al.}{2015}]{chen2015}
{Chen} C.~H.~K.,  {Matteini} L.,  {Burgess} D.,   {Horbury} T.~S.,  2015,
  \mn@doi [MNRAS] {10.1093/mnrasl/slv107}, \href
  {http://adsabs.harvard.edu/abs/2015MNRAS.453L..64C} {453, L64}

\bibitem[\protect\citeauthoryear{Dmitruk, Matthaeus, Milano, Oughton, Zank  \&
  Mullan}{Dmitruk et~al.}{2002}]{dmitruk_etal_2002}
Dmitruk P.,  Matthaeus W.~H.,  Milano L.,  Oughton S.,  Zank G.~P.,   Mullan
  D.,  2002, The Astrophysical Journal, 575, 571

\bibitem[\protect\citeauthoryear{Dmitruk, Matthaeus  \& Oughton}{Dmitruk
  et~al.}{2005}]{dmitruk_etal_2005}
Dmitruk P.,  Matthaeus W.~H.,   Oughton S.,  2005, Physics of Plasmas, 12,
  112304

\bibitem[\protect\citeauthoryear{Einaudi \& Velli}{Einaudi \&
  Velli}{1999}]{einaudi1999}
Einaudi G.,  Velli M.,  1999, Physics of Plasmas, 6, 4146

\bibitem[\protect\citeauthoryear{{Franci}, {Landi}, {Matteini}, {Verdini}  \&
  {Hellinger}}{{Franci} et~al.}{2015}]{franci2015}
{Franci} L.,  {Landi} S.,  {Matteini} L.,  {Verdini} A.,   {Hellinger} P.,
  2015, \mn@doi [\apj] {10.1088/0004-637X/812/1/21}, \href
  {http://adsabs.harvard.edu/abs/2015ApJ...812...21F} {812, 21}

\bibitem[\protect\citeauthoryear{Goldreich \& Sridhar}{Goldreich \&
  Sridhar}{1995}]{goldreich1995}
Goldreich P.,  Sridhar S.,  1995, The Astrophysical Journal, 438, 763

\bibitem[\protect\citeauthoryear{Greco, Matthaeus, Servidio, Chuychai  \&
  Dmitruk}{Greco et~al.}{2009}]{greco_etal_2009}
Greco A.,  Matthaeus W.,  Servidio S.,  Chuychai P.,   Dmitruk P.,  2009, The
  Astrophysical Journal Letters, 691, L111

\bibitem[\protect\citeauthoryear{Howes, Dorland, Cowley, Hammett, Quataert,
  Schekochihin  \& Tatsuno}{Howes et~al.}{2008a}]{howes_etal_2008b}
Howes G.,  Dorland W.,  Cowley S.,  Hammett G.,  Quataert E.,  Schekochihin A.,
    Tatsuno T.,  2008a, Physical Review Letters, 100, 065004

\bibitem[\protect\citeauthoryear{Howes, Cowley, Dorland, Hammett, Quataert  \&
  Schekochihin}{Howes et~al.}{2008b}]{howes_etal_2008}
Howes G.~G.,  Cowley S.~C.,  Dorland W.,  Hammett G.~W.,  Quataert E.,
  Schekochihin A.~A.,  2008b, Journal of Geophysical Research: Space Physics,
  113

\bibitem[\protect\citeauthoryear{Howes, TenBarge, Dorland, Quataert,
  Schekochihin, Numata  \& Tatsuno}{Howes et~al.}{2011}]{howes_etal_2011}
Howes G.~G.,  TenBarge J.~M.,  Dorland W.,  Quataert E.,  Schekochihin A.~A.,
  Numata R.,   Tatsuno T.,  2011, Physical review letters, 107, 035004

\bibitem[\protect\citeauthoryear{{Karimabadi} et~al.,}{{Karimabadi}
  et~al.}{2013}]{karimabadi2013}
{Karimabadi} H.,  et~al., 2013, \mn@doi [Physics of Plasmas]
  {10.1063/1.4773205}, \href
  {http://adsabs.harvard.edu/abs/2013PhPl...20a2303K} {20, 012303}

\bibitem[\protect\citeauthoryear{Kiyani, Osman  \& Chapman}{Kiyani
  et~al.}{2015}]{kiyani_etal_2015}
Kiyani K.~H.,  Osman K.~T.,   Chapman S.~C.,  2015, Philosophical Transactions
  of the Royal Society of London A: Mathematical, Physical and Engineering
  Sciences, 373, 20140155

\bibitem[\protect\citeauthoryear{{Lathrop}}{{Lathrop}}{2005}]{lathrop2005}
{Lathrop} D.,  2005, in Laboratory sodium experiments modeling astrophysical
  and geophysical MHD flows, APS Meeting Abstracts.

\bibitem[\protect\citeauthoryear{Li}{Li}{2008}]{li2008}
Li G.,  2008, The Astrophysical Journal Letters, 672, L65

\bibitem[\protect\citeauthoryear{Mason, Cattaneo  \& Boldyrev}{Mason
  et~al.}{2006}]{mason2006}
Mason J.,  Cattaneo F.,   Boldyrev S.,  2006, \mn@doi [Physical Review Letters]
  {10.1103/PhysRevLett.97.255002}, 97, 255002

\bibitem[\protect\citeauthoryear{{Mason}, {Cattaneo}  \& {Boldyrev}}{{Mason}
  et~al.}{2008}]{mason_cb08}
{Mason} J.,  {Cattaneo} F.,   {Boldyrev} S.,  2008, Physical Review E, 77,
  036403

\bibitem[\protect\citeauthoryear{Mason, Perez, Boldyrev  \& Cattaneo}{Mason
  et~al.}{2012}]{mason2012}
Mason J.,  Perez J.~C.,  Boldyrev S.,   Cattaneo F.,  2012, Physics of Plasmas,
  19, 055902

\bibitem[\protect\citeauthoryear{{Osman}, {Matthaeus}, {Wan}  \&
  {Rappazzo}}{{Osman} et~al.}{2012}]{osman2012}
{Osman} K.~T.,  {Matthaeus} W.~H.,  {Wan} M.,   {Rappazzo} A.~F.,  2012,
  \mn@doi [Physical Review Letters] {10.1103/PhysRevLett.108.261102}, \href
  {http://adsabs.harvard.edu/abs/2012PhRvL.108z1102O} {108, 261102}

\bibitem[\protect\citeauthoryear{Oughton, Matthaeus, Dmitruk, Milano, Zank  \&
  Mullan}{Oughton et~al.}{2001}]{oughton_etal_2001}
Oughton S.,  Matthaeus W.~H.,  Dmitruk P.,  Milano L.,  Zank G.~P.,   Mullan
  D.,  2001, The Astrophysical Journal, 551, 565

\bibitem[\protect\citeauthoryear{Perez \& Boldyrev}{Perez \&
  Boldyrev}{2008}]{perez2008}
Perez J.~C.,  Boldyrev S.,  2008, The Astrophysical Journal Letters, 672, L61

\bibitem[\protect\citeauthoryear{{Perez}, {Mason}, {Boldyrev}  \&
  {Cattaneo}}{{Perez} et~al.}{2012}]{perez_mason2012}
{Perez} J.~C.,  {Mason} J.,  {Boldyrev} S.,   {Cattaneo} F.,  2012, \mn@doi
  [Physical Review X] {10.1103/PhysRevX.2.041005}, \href
  {http://adsabs.harvard.edu/abs/2012PhRvX...2d1005P} {2, 041005}

\bibitem[\protect\citeauthoryear{Rappazzo, Velli, Einaudi  \&
  Dahlburg}{Rappazzo et~al.}{2007}]{rappazzo_etal_2007}
Rappazzo A.,  Velli M.,  Einaudi G.,   Dahlburg R.,  2007, The Astrophysical
  Journal Letters, 657, L47

\bibitem[\protect\citeauthoryear{Rappazzo, Velli, Einaudi  \&
  Dahlburg}{Rappazzo et~al.}{2008}]{rappazzo_etal_2008}
Rappazzo A.,  Velli M.,  Einaudi G.,   Dahlburg R.,  2008, The Astrophysical
  Journal, 677, 1348

\bibitem[\protect\citeauthoryear{Schekochihin, Cowley, Dorland, Hammett, Howes,
  Quataert  \& Tatsuno}{Schekochihin et~al.}{2009}]{schekochihin_etal_2009}
Schekochihin A.,  Cowley S.,  Dorland W.,  Hammett G.,  Howes G.,  Quataert E.,
    Tatsuno T.,  2009, The Astrophysical Journal Supplement Series, 182, 310

\bibitem[\protect\citeauthoryear{TenBarge \& Howes}{TenBarge \&
  Howes}{2013}]{tenbarge_howes_2013}
TenBarge J.,  Howes G.,  2013, The Astrophysical Journal Letters, 771, L27

\bibitem[\protect\citeauthoryear{TenBarge, Howes  \& Dorland}{TenBarge
  et~al.}{2013}]{tenbarge_etal_2013}
TenBarge J.,  Howes G.,   Dorland W.,  2013, The Astrophysical Journal, 774,
  139

\bibitem[\protect\citeauthoryear{{Tessein}, {Ruffolo}, {Matthaeus}  \&
  {Wan}}{{Tessein} et~al.}{2016}]{tessein2016}
{Tessein} J.~A.,  {Ruffolo} D.,  {Matthaeus} W.~H.,   {Wan} M.,  2016, \mn@doi
  [GRL] {10.1002/2016GL068045}, \href
  {http://adsabs.harvard.edu/abs/2016GeoRL..43.3620T} {43, 3620}

\bibitem[\protect\citeauthoryear{Tobias, Cattaneo  \& Boldyrev}{Tobias
  et~al.}{2011}]{tobias2011}
Tobias S.~M.,  Cattaneo F.,   Boldyrev S.,  2011, Ten Chapters in Turbulence

\bibitem[\protect\citeauthoryear{Told, Jenko, TenBarge, Howes  \& Hammett}{Told
  et~al.}{2015}]{told_etal_2015}
Told D.,  Jenko F.,  TenBarge J.,  Howes G.,   Hammett G.,  2015, Physical
  review letters, 115, 025003

\bibitem[\protect\citeauthoryear{Wan, Rappazzo, Matthaeus, Servidio  \&
  Oughton}{Wan et~al.}{2014}]{wan_etal_2014}
Wan M.,  Rappazzo A.~F.,  Matthaeus W.~H.,  Servidio S.,   Oughton S.,  2014,
  The Astrophysical Journal, 797, 63

\bibitem[\protect\citeauthoryear{{Wu}, {Wan}, {Matthaeus}, {Shay}  \&
  {Swisdak}}{{Wu} et~al.}{2013}]{wu2013}
{Wu} P.,  {Wan} M.,  {Matthaeus} W.~H.,  {Shay} M.~A.,   {Swisdak} M.,  2013,
  \mn@doi [Physical Review Letters] {10.1103/PhysRevLett.111.121105}, \href
  {http://adsabs.harvard.edu/abs/2013PhRvL.111l1105W} {111, 121105}

\bibitem[\protect\citeauthoryear{Zhdankin, Boldyrev, Mason  \& Perez}{Zhdankin
  et~al.}{2012}]{zhdankin_etal_2012}
Zhdankin V.,  Boldyrev S.,  Mason J.,   Perez J.~C.,  2012, Physical Review
  Letters, 108, 175004

\bibitem[\protect\citeauthoryear{Zhdankin, Boldyrev  \& Uzdensky}{Zhdankin
  et~al.}{2016}]{zhdankin_etal_2016b}
Zhdankin V.,  Boldyrev S.,   Uzdensky D.~A.,  2016, Physics of Plasmas, 23,
  055705

\makeatother
\end{thebibliography}


\bsp	
\label{lastpage}
\end{document}